\documentclass[prb,twocolumn,showpacs,amsmath,amssymb,superscriptaddress,longbibliography]{revtex4-1}

\usepackage{graphicx}
\usepackage{ifthen}
\usepackage{color}
\usepackage{bm}
\usepackage{braket}
\usepackage{multirow}
\usepackage{hyperref}
\hypersetup{backref=true,
 pdfnewwindow=true, colorlinks=true,
 linkcolor=blue, anchorcolor=blue,
 citecolor=blue, filecolor=blue,
 menucolor=blue, urlcolor=blue}

\begin{document}

\title{Spin contribution to the inverse Faraday effect of non-magnetic metals}
\author{Shashi B. Mishra}
\affiliation{Mechanical Engineering, University of California, Riverside, California 92521, USA.}
\author{Sinisa Coh}
\affiliation{Mechanical Engineering, University of California, Riverside, California 92521, USA.}
\affiliation{Materials Science and Engineering, University of California, Riverside, California 92521, USA.}
\date{\today}

\begin{abstract}
We formulate the spin contribution to the inverse Faraday effect of non-magnetic metals.  We deal with the role of the inversion symmetry, which forces all electronic bands to be at least twice degenerate at every point in the Brillouin zone.  We show both analytically and numerically that our formulation of the inverse Faraday effect is invariant under unitary rotation within the doubly degenerate set of bands.  In addition, we show the importance of resonance-like features in the band structure for the inverse Faraday effect.  Our first-principles computed spin component of the inverse Faraday effect in a simple metal such as Au is reminiscent of its optical absorption, with a characteristic d--s resonance in the optical spectrum.
\end{abstract}

\maketitle

\section{\label{sec:intro}Introduction}

The inverse Faraday effect (IFE) is a phenomenon in which circularly polarized (CP) light acts as an effective magnetic field that induces static magnetization in a material. It was first predicted phenomenologically in the 1960s.\cite{Pitaevskii1960}  It was observed experimentally by van der Ziel \textit{et al.}\cite{van1965} in CaF$_2$ doped with magnetic impurities.  The work of Pershan \textit{et al.}~\cite{Pershan1966} provided a more detailed analysis, based on a quantum-mechanical model of a localized magnetic impurity. Interest in IFE has recently been renewed after the experimental demonstration of the control of spin dynamics in magnets by Kimel \textit{et al}.\cite{Kimel2005} Optical magnetic switching has also been reported in ferrimagnetic GdFeCo,\cite{Stanciu2007} TbCo,\cite{Mangin_2012} ferromagnetic Co/Pt bilayer,\cite{Lambert2014} as well as other materials.\cite{Mangin2014}  More recent reports assign the control of magnetism in these materials to thermal effects that do not involve IFE.\cite{Ostler2012,Gorchon2016,Gorchon-II-2016} Nevertheless, IFE has been emerging as one of the potential ways for ultrafast data processing.\cite{RevModPhys2010_Rasing} Understanding the theoretical mechanism behind IFE is relevant for further progress in the field of ultrafast magnetism. Theoretically, IFE has been formulated for graphene, as well as Weyl-semimetals,\cite{Tokman_2020,gao2020topological} Rashba metals,\cite{Tanaka_2020} Mott insulators,\cite{Banerjee2022} ferromagnets\cite{Berritta2016,Scheid2019,Freimuth2016} and non-magnetic nanomaterials such as gold nanoparticles.\cite{Cheng2020,Gu_2010,Hurst2018,Smolyaninov2005} The IFE theory based on first principles is formulated both considering only spin~\cite{Freimuth2016,Scheid2019} and both spin and local part of orbital magnetization.\cite{Battiato2014,Berritta2016} Some of the IFE theories are semiclassical and take into account the hydrodynamic description of the free electron gas.\cite{HERTEL2006L1,Nadarajah2017,Hurst2018,Sinha-Roy2020}  Earlier work by Wagniere\cite{Wagniere1989} and Volkov\cite{Volkov2002} discussed the symmetry property of IFE.

In this work, we revisit the first-principles-based IFE theory for nonmagnetic metals. In particular, we focus on the role of the inversion symmetry ($\mathcal {P}$) present in the bulk crystal structure of most metals, including simple metals such as Cu or Au. Nonmagnetic metals with inversion symmetry contain in their magnetic point group the $\cal{PT}$ operation, which is a combination of time-reversal ($\mathcal{T}$) and spatial inversion ($\mathcal{P}$) operations. Due to the presence of $\cal{PT}$ symmetry, all electron bands at all points in the Brillouin zone are at least twice degenerate.\cite{Elliott1954} While in many physical situations the presence of this band degeneracy does not cause difficulties in computing various physical properties, the case with IFE computation is somewhat more involved. In this work, we show how to deal with the presence of $\cal{PT}$ symmetry in the calculation of IFE so that the final result does not depend on the arbitrary unitary mixture of the states in the doubly degenerate subspace. Our approach is easily applicable in the first-principles context, and we demonstrate it in the case of the calculation in bulk Au.  Finally, we also discuss the role of band resonance in the computed inverse Faraday effect.

The remainder of the paper is organized as follows. In Sec.~\ref{sec:results}, we derive our theory and demonstrate its invariance under unitary band transformation, both analytically and numerically using first-principles calculations on bulk gold. We discuss our results in Sec.~\ref{sec:discussion} and summarize in Sec.~\ref{sec:summary}. Additional details are provided in the Appendices. 

\section{\label{sec:results}Results}

When light is incident on a metal, the electron is excited to a higher energy state. If the electron moves to a higher energy state within the same energy band, the transition is known as intraband. When the electron is excited to a different energy band, it is known as an interband transition. In the case of the IFE, we are interested in the magnetic moment induced via the interaction of electrons with the CP light. Therefore, IFE will, in general, have contributions from either interband or intraband transitions. Furthermore, the magnetic moment that is induced by the CP light can originate either from spin or orbital degrees of freedom.   

\begin{figure}[!t]
    \centering
    \includegraphics[width=3.4in]{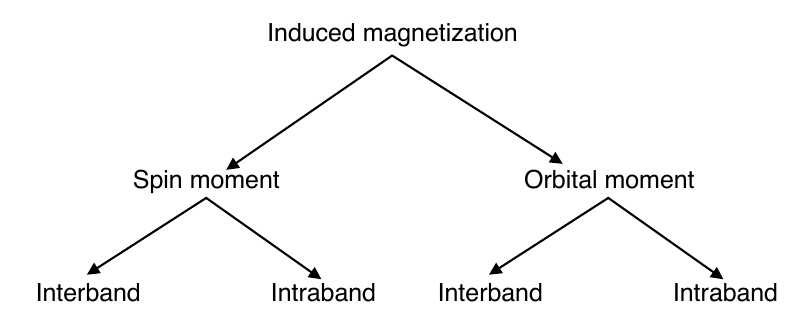}
    \caption{\label{fig:schematic_magn} Inverse Faraday effect contributions in a metal.}
\end{figure}

Therefore, in total, the complete theory of IFE in metal would have to consist of four components: interband-spin, intraband-spin, interband-orbital, and intraband-orbital, as sketched in Fig.~\ref{fig:schematic_magn}.

\subsection{Working assumptions}

In this work, we focus on the spin component of the IFE. The reason for focusing on the spin component is two-fold. First, the formulation of the spin part of the IFE is significantly more straightforward than the formulation of the orbital moment. In fact, the modern theory of orbital magnetization in an infinite bulk periodic solid has been developed only relatively recently,\cite{Thonhauser_2005,Xiao_2005,Ceresoli_2006,Shi_2007}
while the theory of induced orbital magnetization in an electric field (so far, static field, in an insulator) is even more recent.\cite{Andrei_2011} Therefore, thus far the orbital part of the inverse Faraday effect has been included only on the local level,\cite{Battiato2014,Berritta2016} or within the semi-classical approach.\cite{Hurst2018}

The intraband orbital contribution seems especially problematic within the modern theory of orbital moment, as such theories have so far been formulated only for electronic states in a periodic solid with a well-defined crystal momentum $\bm k$. On the other hand, within the Lindhard-like approach, one needs to work with states that are linear combination of $\bm k$ and ${\bm k} + {\bm q}$ where $\bm q$ is the wavevector of light.\cite{Lindhard1954} Therefore, we leave the discussion of the orbital contribution to the IFE for future work.

The second reason for focusing on the spin part of IFE is that in this work we are mainly interested in the role of the electron band degeneracy in the formulation of IFE for $\cal{PT}$-symmetric metals, as well as the role of resonance. These two issues, degeneracy and resonance, are already present in the spin part of the IFE, so they can be addressed without considering the orbital part. 

Although the spin part of IFE has, in principle, both interband and intraband contributions, we will show that, in the case of nonmagnetic material, the intraband-spin part of IFE is negligible at optical frequencies. Therefore, in the end, it will suffice to focus here on only one of the four components of IFE, namely, the interband-spin component.

We formulate the theory of spin IFE using standard adiabatic time-dependent perturbation theory. The perturbation is turned on infinitely slowly, which ensures that there are no discontinuities in theory. We refer the reader to Ref.~\onlinecite{Langhoff1972} for a detailed review of the adiabatic time-dependent perturbation theory. Since the perturbation of the incoming light is turned on infinitesimally slowly, our theory does not have a transient behavior that originates from a suddenly turned-on perturbation. Instead, when light with frequency $\omega$ has been impinging on the solid for a long time, we find that there are only two contributions to the induced magnetization. One is constant in time, while the other oscillates with a frequency of $2\omega$. We show later in Sec.~\ref{sec:non-doubly-res} that the $2 \omega$ term is typically smaller in magnitude than the constant term, which is why we focus on the term that is constant in time.

In this work, we treat the role of disorder by simply assuming a constant lifetime of carriers that is independent of both the band index and the $\bm k$-point.

\subsection{Derivation}\label{sec:derivation}

Now, we are ready to compute the spin component of the inverse Faraday effect. We start with the time-dependent Schr\"{o}dinger-like equation within the independent particle approximation,
\begin{align}\label{eq:full_sch}
    H(\lambda, t) \Psi_i(\lambda,t) = i\hbar\frac{\partial \Psi_i(\lambda,t)}{\partial t}.
\end{align}
For now, we use a generic index $i$ to distinguish the states $\Psi_i$. These states are single Slater determinants corresponding to fully occupied single-electron states below the Fermi level $E_{\rm F}$. Hamiltonian $H$ appearing in Eq.~\ref{eq:full_sch} is also a $2 \times 2$ matrix in spin indices, as it includes relativistic effects such as spin-orbit interaction. However, for brevity, here we suppress the spinor indices in $H$ and $\Psi$. 

The Hamiltonian $H(\lambda, t)$ consists of the unperturbed part $H_0$ and the perturbation $V$ applied with frequency $\omega > 0$,
\begin{align}\label{eq:full_ham}
H = H_0 + \lambda e^{\delta t} \left( V e^{i \omega t} + V^{\dagger} e^{-i \omega t} \right).
\end{align}
For a moment, we keep the perturbation $V$ generic and later replace it with the interaction of the electron with the external electric field (see Appendix~\ref{app:light-electron} for more details). We will assume that the electric field of light is polarized in the $x$--$y$ plane with a definite helicity (as specified in Appendix~\ref{app:light-electron}).  The induced magnetic moment in a cubic material must then be along the $z$-axis. The strength of the interaction is parameterized with the dimensionless parameter $\lambda$. In the limit $\delta \rightarrow 0^+$, the switching function $e^{\delta t}$ is responsible for turning on the perturbation $V$ infinitesimally slowly.

We will solve the time-dependent equation in terms of the solutions $\Phi_i$ to the unperturbed time-independent  Hamiltonian,
\begin{align}\label{eq:unpert_ham}
    H_0 \Phi_i = E_i \Phi_i.
\end{align}
Following Ref.~\onlinecite{Langhoff1972} and using a specific formulation from Ref.~\onlinecite{Bhattacharyya1986}, the solution of the time-dependent problem as a series expansion in $\lambda$ is given by,
\begin{eqnarray}\label{eq:full_sol}
\Psi_i = N_i e^{-i \alpha_i t} \left( \Phi_i+\sum_j^{j \neq i} B_j \Phi_j e^{i \alpha_{ij} t} \right).
\end{eqnarray}
The coefficients $B_j$ are expanded to powers of $\lambda$,
\begin{align}\label{eq:B_expansion}
B_j = B_j^{(0)} + \lambda B_j^{(1)} + \lambda^2 B_j^{(2)} + \ldots
\end{align}
as detailed in Appendix~\ref{app:perturbation}. We choose $B_j^{(0)}=0$ for all $j$.

In the lowest order in perturbation theory, $\hbar \alpha_{ij} = E_i - E_j$.  For completeness, we provide the Taylor expansion of $\alpha_{ij}$ to second order in $\lambda$ in Appendix~\ref{app:perturbation}. The sum in Eq.~\ref{eq:full_sol} is over all states $j$ that do not equal the unperturbed state $i$. 

We computed $|N_i|^2$ by imposing the normalization condition, $\braket{\Psi_i | \Psi_i}=1$. From the orthonormality of the unperturbed states $\braket{\Phi_i | \Phi_j} = \delta_{ij}$, it trivially follows that
\begin{align}\label{eq:normalization}
| N_i |^2 = 1 + {\cal O} (\lambda^2).
\end{align}

Now we compute the spin magnetic moment per unit cell in the perturbed state $\Psi_i$. This is simply given as the expectation value of the spin moment operator,
\begin{align}\label{eq:mag_def}
M_i^{\rm IFE} (t) = \braket{ \Psi_i | M^{\rm spin} | \Psi_i}.
\end{align}
The total spin-magnetic moment $M_i^{\rm IFE}$ must vanish at order $\lambda^0$, as we assume that the ground state is non-magnetic. The linear term $M_i^{\rm IFE} \sim \lambda^1$ must vanish by symmetry.  Therefore, in the lowest order $M_i^{\rm IFE}$ scales as $\lambda^2$. Using Eqs.~\ref{eq:full_sol} and \ref{eq:B_expansion} in Eq.~\ref{eq:mag_def} gives us the following for the magnetic moment,
\begin{align}\label{eq:msecond}
M_i^{\rm IFE} (t)  = 2 {\rm Re} \left[ \sum_j^{j \neq i}B_j^{(2)} 
e^{i \alpha_{ij} t} \bra{\Phi_i}M^{\rm spin}\ket{\Phi_j}\right] +
\notag \\ 
+ 
\sum_j^{j \neq i} \left[ B_j^{(1)} \right]^* e^{-i \alpha_{ij} t}
\sum_l^{l \neq i } B_{l}^{(1)} e^{i \alpha_{il} t}
\bra{\Phi_j}M^{\rm spin}\ket{\Phi_{l}}.
\end{align}
Here, we collected all terms in the expansion that scale as $\lambda^2$ and neglected all higher orders of $\lambda$. From now on, we set $\lambda=1$ for simplicity. 

In deriving Eq.~\ref{eq:msecond} we had to take into account that the $\lambda^2$ contribution of $|N_i|^2$ from Eq.~\ref{eq:normalization} is now weighted by the expectation of the spin magnetic moment in the unperturbed ground state, $\braket{ \Phi_i | M^{\rm spin} | \Phi_i}$.  Since the expectation value of total spin moment is zero in a non-magnetic system, we conclude that the normalization term $|N_i|^2$ does not contribute to $M_i^{\rm IFE} (t)$ in the $\lambda^2$ order.

Inserting directly $B_j^{(1)}$ and $B_j^{(2)}$ from Eqs.~\ref{eq:B1} and \ref{eq:B2} into our expression for $M_i^{\rm IFE} (t)$ (Eq.~\ref{eq:msecond}) would give us both constant in time contributions to $M_i^{\rm IFE}$, as well as those that oscillate with a frequency of $2 \omega$. As shown in Sec.~\ref{sec:non-doubly-res}, the $2 \omega$ contribution is smaller in magnitude than the constant contribution. Therefore, from now on, we focus on the constant time-independent contribution and we will denote the corresponding spin expectation value simply as $M_i^{\rm IFE}$ without explicit time dependence. Inserting Eqs.~\ref{eq:B1} and \ref{eq:B2} into Eq.~\ref{eq:msecond} and rearranging the terms, we find the following expression for the constant time-independent contribution to IFE,
\begin{align}
\label{eq:ife_Bj}
M_i^{\rm IFE} = & 
\sum_j^{j \neq i} \sum_l^{l \neq i } 
\left[
\frac{
\braket{ \Phi_i | V| \Phi_j}
\braket{ \Phi_j | M^{\rm spin} | \Phi_l}
\braket{ \Phi_l | V^\dagger | \Phi_i}
}
{
(E_j - E_i - \hbar \omega - i \eta)
(E_l - E_i - \hbar \omega + i \eta)
}
\right.
\nonumber 
\\ 
+
&
\frac{
\braket{ \Phi_i | V^\dagger| \Phi_j}
\braket{ \Phi_j | M^{\rm spin} | \Phi_l}
\braket{ \Phi_l | V | \Phi_i}
}
{
(E_j - E_i + \hbar \omega - i \eta)
(E_l - E_i + \hbar \omega + i \eta)
}
\nonumber 
\\ 
+  
&
2 {\rm Re} 
\frac{
\braket{ \Phi_i | M^{\rm spin} | \Phi_j}
\braket{ \Phi_j | V | \Phi_l}
\braket{ \Phi_l | V^\dagger | \Phi_i}
}
{
(E_j - E_i + 2 i \eta)
(E_l - E_i - \hbar \omega + i \eta)
} 
\nonumber 
\\ 
+
&
2 {\rm Re} 
\left.
\frac{
\braket{ \Phi_i | M^{\rm spin} | \Phi_j}
\braket{ \Phi_j | V^\dagger | \Phi_l}
\braket{ \Phi_l | V | \Phi_i}
}{
(E_j - E_i + 2 i \eta)
(E_l - E_i + \hbar \omega + i \eta)
}
\right].
\end{align}
Although the perturbation theory formulation from Ref.~\onlinecite{Langhoff1972, Bhattacharyya1986} is in principle without divergences when using the recursive formulation of $\alpha_{ji}$ from Eq.~\ref{eq:alpha_nm}, such a formulation is numerically intensive and excludes sources of electron level broadening that are not included in $V$, such as phonons or defects. Therefore, as discussed in the Appendix~\ref{app:perturbation}, we assumed here that electronic states have a constant lifetime proportional to $\eta^{-1}$. In addition, we replaced here $\hbar \alpha_{ij}$ with its lowest order expansion, $E_i - E_j$.

As we can see from Eq.~\ref{eq:ife_Bj}, there are four groups of contributions to $M_i^{\rm IFE}$ that are constant over time. However, we expect that only some of these will dominate. In particular, we expect that the first term in the expression will dominate whenever the denominator is close to zero.  For example, such a resonant condition can occur whenever there is a state $j$ for which $E_j - E_i$ is close to $\hbar \omega$.  If in addition, we consider a state $l$ such that $E_j = E_l$, then we expect a doubly-resonant condition, as both denominators in the first term could then simultaneously be close to zero. (Here $E_j$ and $E_l$ are strictly larger than $E_i$ as $i$ is the ground state.) Later, in Sec.~\ref{sec:non-doubly-res}, we show with an explicit numerical calculation that, in fact, these doubly resonant terms dominate the IFE response in bulk Au.

From now on we will explicitly separate out the doubly-resonant part from the first term in Eq.~\ref{eq:ife_Bj}. We are then left with the doubly-resonant contribution, and four groups of non-doubly resonant terms,
\begin{align}\label{eq:m_resonant}
M_i^{\rm IFE} = & \sum_j^{j \neq i} \overset{E_l = E_j}{\sum_l^{l \neq i}} 
\frac{
\braket{\Phi_i | V | \Phi_j} 
\braket{\Phi_j | M^{\rm spin} | \Phi_l}
\braket{\Phi_l | V^{\dagger} | \Phi_i} }
{(E_j - E_i - \hbar \omega)^2 + \eta^2} 
\notag \\
& + (\textrm{non-doubly-resonant terms}).
\end{align}
Here the second sum over $l$ is done only over states $l$ that have the same energy as the state $j$ appearing in the first sum.

Up until now, the indices $i$, $j$, and $l$ were simply labeling many-electron states in the system within the independent-electron approximation. From now on, we will introduce changes to this notation. First, we will do the following replacements,
\begin{align}
&\ket{\Phi_i}  \longrightarrow \ket{\rm GS}, \label{eq:2ndquant_i} \\
&\ket{\Phi_j}  \longrightarrow c^{\dagger}_{m  M  {\bm K}} c^{\phantom{\dagger}}_{n  N  {\bm k}} \ket{\rm GS}, \label{eq:2ndquant_j} \\
&\ket{\Phi_l}  \longrightarrow c^{\dagger}_{m' M' {\bm K}} c^{\phantom{\dagger}}_{n' N' {\bm k}} \ket{\rm GS}. \label{eq:2ndquant_l}
\end{align}
Here $\ket{\rm GS}$ is the ground state in which every single-particle orbital with energy below the Fermi level $E_{\rm F}$ is occupied. The electron destruction operator $c_{n N {\bm k}}$ corresponds to the single-particle Bloch orbitals $\ket{\phi_{nN {\bm k}}}$. Here, $\bm k$ is the electron crystal momentum. In what follows, we will usually not write crystal momenta explicitly, and we are going to assume that for the interband-transitions ${\bm k}={\bm K}$. In the case of the intraband-transitions, we assume that the $\bm k$ and $\bm K$ differ by the wavevector of the incoming light, and then we work in the ${\bm k} \rightarrow {\bm K}$ limit. In addition to crystal momentum, we also label different electron bands with a pair of indices $(n, N)$, since each electronic band is at least twice degenerate due to the $\cal{PT}$ symmetry. In particular, we will label band doublets with index $n$ and distinguish individual states in the doublet with an additional index $N$. For each $n$, we will label one state with $N=1$ and another with $N=2$. As discussed in Sec.~\ref{sec:gauge_inv}, there is some freedom in choosing single-particle orbitals that correspond to $N=1$ or $N=2$. (We note that without spin-orbit interaction, $N=1$ and $N=2$ could simply correspond to two different eigenstates of the electron spin operator $S_z$. Therefore, one could choose $N=1$ to correspond to the state with spin pointing along the $+\hat{\bm z}$ direction and $N=2$ with spin along $-\hat{\bm z}$. However, when the spin-orbit interaction is included in our calculation, the states no longer have a well-defined projection of the spin $S_z$.)

The earlier requirement in Eq.~\ref{eq:m_resonant} that $i \neq j$ is now in the context of Eqs.~\ref{eq:2ndquant_i} and \ref{eq:2ndquant_j} converted to the requirement that the labels in the subscript of $c^{\phantom{\dagger}}_{n  N  {\bm k}}$ correspond to the state with energy below $E_{\rm F}$, while the labels in $c^{\dagger}_{m  M  {\bm k}}$ correspond to the state with energy above $E_{\rm F}$.  A similar requirement follows from $i \neq l$.  With this replacement, and using our notation for a doublet, converting the sum over states integral over the Brillouin zone, and using the second-quantized form of the $V$ and $M^{\rm spin}$ operators, we get the following for the induced magnetic moment $M$ per unit cell,
\begin{widetext}
\begin{align}
M^{\rm IFE} = & 
M^{\rm IFE}_{\rm elec}
-
M^{\rm IFE}_{\rm hole}
+
M^{\rm IFE}_{\rm ndr}
\label{eq:ife_all_three}
\\
M^{\rm IFE}_{\rm elec} = & 
\int_{\rm BZ} 
\frac{d^3k}{(2\pi)^3} 
\sum_n^{\rm occ} 
\sum_{N=1}^2
\sum_m^{\rm emp}
\sum_{M=1}^2
\sum_{M'=1}^2
\frac{
\braket{\phi_{nN} | V | \phi_{mM}} 
\braket{\phi_{mM} | M^{\rm spin} | \phi_{mM'}}
\braket{\phi_{mM'} | V^{\dagger} | \phi_{nN}} }
{(E_m - E_n - \hbar \omega)^2 + \eta^2}
\label{eq:ife_elec}
\\
M^{\rm IFE}_{\rm hole} = & 
\int_{\rm BZ} 
\frac{d^3k}{(2\pi)^3} 
\sum_n^{\rm occ} 
\sum_{N=1}^2
\sum_{N'=1}^2
\sum_m^{\rm emp}
\sum_{M=1}^2
\frac{
\braket{\phi_{nN} | V | \phi_{mM}} 
\braket{\phi_{mM} |  V^{\dagger} | \phi_{nN'}}
\braket{\phi_{nN'} | M^{\rm spin} | \phi_{nN}} }
{(E_m - E_n - \hbar \omega)^2 + \eta^2}
\label{eq:ife_hole}
\end{align}
The terms $M^{\rm IFE}_{\rm elec}$ and $M^{\rm IFE}_{\rm hole}$ can be doubly-resonant and they originate from the first term in Eq.~\ref{eq:m_resonant}.  The remaining, non-doubly resonant, term $M^{\rm IFE}_{\rm ndr}$ is given in the Appendix~\ref{app:nonresonant}.  We compute the needed matrix elements as,
\begin{align}
\label{eq:Moper}
&\braket{\phi_{mM} | M^{\rm spin} | \phi_{mM'}} 
=
2 \frac{e}{2 m_{\rm e}} 
\braket{\phi_{mM} | S_z | \phi_{mM'}},
\\
\label{eq:Voper}
&\braket{\phi_{nN} | V | \phi_{mM}} 
=
\frac{e}{2}\sqrt{\frac{I}{\epsilon_0 c}} 
\frac{E_n - E_m}{\hbar \omega}
\left( A_{nNmM}^x + i A_{nNmM}^y \right) \textrm{\quad (interband, $n\neq m$)}.
\end{align}
\end{widetext}
and we define the Berry connection $A_{nNmM}^{\alpha}$ as,
\begin{align}
\label{eq:berry_conn}
A_{nNmM}^{\alpha} = \braket{u_{nN}| i \partial_{k_{\alpha}} | u_{mM}}.
\end{align}
The expression for the intraband optical transition is given in Eq.~\ref{eq:intraband}.

In the above expressions, we don't include doublet indices $N$, $N'$, $M$, and $M'$ in the eigenenergy $E_{n}$ as, by definition, states in the doublet have the same energy, so $E_n=E_{nN}$. The charge of the electron is $e$ and its mass is $m_{\rm e}$. The spin angular momentum operator is $S_z$ with the expectation value of $\pm \hbar/2$ for a fully spin-polarized electron.  The electron orbitals are normalized to unity in a single unit cell. As discussed earlier, here we assumed that the electric field of light is polarized in the $x$--$y$ plane and that the induced magnetic moment is along the $z$-axis. In Eq.~\ref{eq:Voper}, we denoted with $I$ the intensity of the incoming light, $c$ is the speed of light, and $\epsilon_0$ is the permittivity of the free space. The cell-periodic part of the Bloch state $\ket{\phi_{nN}}$ we denoted as $\ket{u_{nN}}$. The derivative with respect to the $\alpha$, Cartesian component of the electron momentum $k_{\alpha}$, we denote as $\partial_{k_{\alpha}}$.

The sums over $N$, $N'$, $M$, and $M'$ in Eqs.~\ref{eq:ife_elec}, \ref{eq:ife_hole}, and \ref{eq:ndr-a}--\ref{eq:ndr-h} are done over the degenerate subspace.  In these expressions, we assumed that there are two degenerate bands at each $\bm k$-point, but the generalization to a different number of degenerate bands, including bands that are non-degenerate, is trivial.

\subsection{\label{sec:gauge_inv} Degenerate band gauge invariance}

Our expressions Eq.~\ref{eq:ife_elec} and \ref{eq:ife_hole}, for the induced magnetic moment, are written in terms of the single-particle orbitals $\phi_{nN}$. These are eigenstates used to construct Slater determinants of the unperturbed Hamiltonian given in Eq.~\ref{eq:unpert_ham}. As discussed earlier, the choice of orbitals $\phi_{nN}$ is not unique. In the presence of $\cal{PT}$ symmetry, all bands are doubly degenerate. Therefore, for each doublet $n$, we could have chosen, as our perturbation basis, any linear combination of states $\phi_{nN=1}$ and $\phi_{nN=2}$ within the doublet. Formally, we could have rotated each doublet $n$ at each ${\bm k}$-point using an arbitrary $2 \times  2$ unitary matrix $U_{n {\bm k}}$ as follows,
\begin{align}
\label{eq:gen_gauge}
\phi_{n N} \longrightarrow \sum_{P=1}^2  U_{n {\bm k}}^{P N} \phi_{nP}.
\end{align}
Now we will confirm that our Eq.~\ref{eq:ife_elec} is indeed invariant under the transformation in Eq.~\ref{eq:gen_gauge}. The transformation in Eq.~\ref{eq:gen_gauge} does not mix states of different energy, as such a transformation would result in states that are not eigenstates of the unperturbed Hamiltonian given in Eq.~\ref{eq:unpert_ham}. Therefore, instead of focusing on the entire Eq.~\ref{eq:ife_elec}, it is enough to show the degenerate band gauge invariance of the numerator for fixed doublets $n$ and $m$,
\begin{widetext}
\begin{align}
\label{eq:numer}
g_{nm}^{\rm elec} = 
\sum_{N=1}^2
\sum_{M=1}^2
\sum_{M'=1}^2
\braket{\phi_{nN} | V | \phi_{mM}} 
\braket{\phi_{mM} | M^{\rm spin} | \phi_{mM'}}
\braket{\phi_{mM'} | V^{\dagger} | \phi_{nN}} 
 \textrm{ \ \ with \ \ } (n \neq m).
\end{align}
\end{widetext}
Furthermore, for simplicity, we will focus here on the degenerate band gauge invariance of only $M^{\rm IFE}_{\rm elec}$ from Eq.~\ref{eq:ife_elec}. The demonstration of degenerate band gauge invariance of $M^{\rm IFE}_{\rm hole}$ and $M^{\rm IFE}_{\rm ndr}$ proceeds in a similar fashion. Here, we are allowed to consider only the case $n \neq m$, since $M^{\rm IFE}_{\rm elec}$ is written as the sum of $n$ in the occupied band and $m$ in the empty band. Clearly, with the definition of $g_{nm}^{\rm elec}$, the Eq.~\ref{eq:ife_elec} can be rewritten as 
\begin{align}\label{eq:ife_simpler}
M^{\rm IFE}_{\rm elec} &= \int_{\rm BZ} 
\frac{d^3k}{(2\pi)^3} 
\sum_n^{\rm occ} 
\sum_m^{\rm emp}
\frac{
g_{nm}^{\rm elec} }
{(E_m - E_n - \hbar \omega)^2 + \eta^2}.
\end{align}
Since states transform according to Eq.~\ref{eq:gen_gauge}, the spin magnetic moment matrix elements $\braket{\phi_{mM} | M^{\rm spin} | \phi_{mM'}}$ transform as,
\begin{align}
\label{eq:transf_M}
\braket{\phi_{mM} | M^{\rm spin} | \phi_{mM'}}
\longrightarrow 
U_{m}^{\dagger M P}
\braket{\phi_{mP} | M^{\rm spin} | \phi_{mP'}}
U_{m}^{P' M'}.
\end{align}
Here we implicitly assume the sum over repeated indices $P$ and $P'$.

The transformation of the matrix element of $V$ at first seems somewhat more involved than that of $M^{\rm spin}$, as it contains the derivative with respect to $\bm k$. Therefore, this derivative must also act on our unitary rotation matrix, $U_{n {\bm k}}$ as it is generally $\bm k$-dependent.  Therefore, following the definition of Berry connection in Eq.~\ref{eq:berry_conn}, it transforms as,
\begin{align}
A_{nNmM}^{\alpha} 
\longrightarrow
U_{n {\bm k}}^{\dagger N P}
\bra{\phi_{nP}} i \partial_{k_{\alpha}} \left( \ket{ \phi_{m R}}
U_{m {\bm k}}^{R M} \right).
\end{align}
It is easy to see that in our case it is enough to consider only the term where the derivative $\partial_{k_{\alpha}}$ acts on the ket,
\begin{align}
\label{eq:transf_A}
A_{nNmM}^{\alpha} 
\longrightarrow
U_{n {\bm k}}^{\dagger N P}
A_{nPmR}^{\alpha} 
U_{m {\bm k}}^{R M}, \textrm{ if } n \neq m.
\end{align}
The remaining term, where derivative acts on the matrix $U_{m {\bm k}}^{R M}$ vanishes, as it is proportional to $\braket{\phi_{nP} | \phi_{m R}}$ which is identically zero due to our assumption that $n \neq m$.\footnote{We note here that when considering the intraband contribution to the IFE we would need to include the $n=m$ case.  However, in this case, the optical transition matrix element has a particularly simple form, given by Eq.~\ref{eq:intraband}. Clearly, such a matrix element transforms as a scalar, as it is proportional to the Kronecker delta in the doublet indices.} (We note that the second term is generally non-zero for a multi-band gauge transformation that can mix one-particle states of different energy. We refer to Ref.~\onlinecite{Marzari1997,Wang2006} for more details on such transformations.) It follows trivially that the matrix element of $V$ then also transforms as a simple matrix,
\begin{align}
\label{eq:transf_V}
\braket{\phi_{nN} | V | \phi_{mM}}
\longrightarrow 
U_{n}^{\dagger N P}
\braket{\phi_{nP} | V | \phi_{mR}}
U_{m}^{R M},  \textrm{ if } n \neq m.
\end{align}

Since both Eqs.~\ref{eq:transf_M} and \ref{eq:transf_V} are simple unitary matrix transformations, it follows that $g_{nm}^{\rm elec}$ transforms as a trace of the product of three matrices. The trace of a matrix is invariant under unitary transformation, so $g_{nm}^{\rm elec}$ is also invariant,
\begin{align}
\label{eq:transf_g}
g_{nm}^{\rm elec}
\longrightarrow
g_{nm}^{\rm elec}.
\end{align}
From Eq.~\ref{eq:ife_simpler}, it now follows that $M^{\rm IFE}_{\rm elec}$ itself is also degenerate band gauge invariant under transformation in Eq.~\ref{eq:gen_gauge}
\begin{align}
\label{eq:transf_ife}
M^{\rm IFE}_{\rm elec}
\longrightarrow
M^{\rm IFE}_{\rm elec}.
\end{align}
This concludes the proof of the degenerate band gauge invariance of $M^{\rm IFE}_{\rm elec}$. Since the numerators of the expression for $M^{\rm IFE}_{\rm hole}$ and $M^{\rm IFE}_{\rm ndr}$ are very similar to those of $M^{\rm IFE}_{\rm elec}$, their degenerate band gauge-invariance can be demonstrated analogously following the same approach.

\subsection{Comparison to degenerate band gauge invariance in previous works}
\label{sec:previous}

Now we compare the IFE given by our Eqs.~\ref{eq:ife_elec} and \ref{eq:ife_hole} with that of Refs.~\onlinecite{Scheid2019,Battiato2014,Berritta2016}. The expressions in these previous works can be rewritten in a form similar to that of Eq.~\ref{eq:ife_simpler} but with a different choice of the numerator $g_{nm}^{\rm elec}$ (the energy denominator in the case of Refs.~\onlinecite{Battiato2014} and \onlinecite{Berritta2016} is different from our Eqs.~\ref{eq:ife_elec} and \ref{eq:ife_hole}, as discussed in Sec.~\ref{sec:num_comparison}, but this difference does not affect the degenerate band gauge invariance). 

We start by discussing first Ref.~\onlinecite{Scheid2019}. This work reports on the calculation of the IFE-like response in materials without $\cal{PT}$ symmetry, such as ferromagnetic iron. The numerator appearing in Ref.~\onlinecite{Scheid2019} is of the following form,
\begin{align}
\label{eq:numer_scheid} 
\braket{\phi_{n} | V | \phi_{m}} 
\braket{\phi_{m} | M^{\rm spin} | \phi_{m}}
\braket{\phi_{m} | V^{\dagger} | \phi_{n}}.
\end{align}
Clearly, the difference between our Eq.~\ref{eq:numer} and Eq.~\ref{eq:numer_scheid} is the lack of sum over the indices that distinguish the states in the doublet ($N$, $M$, and $M'$). However, this is to be expected, as materials considered in Ref.~\onlinecite{Scheid2019} do not have $\cal{PT}$ symmetry (and thus don't have doubly degenerate band structure at all k-points) so the indices $n$ and $m$ now run over all electronic states at a given $\bm k$-point. Eq.~\ref{eq:numer_scheid} is trivially degenerate band gauge-invariant under the following transformation,
\begin{align}
\label{eq:gen_gauge_nondeg}
\phi_{n {\bm k}} \longrightarrow  U_{n {\bm k}} \phi_{n {\bm k}}
\end{align}
as $U_{n {\bm k}}$ is no longer a matrix, but simply a complex number with norm 1. Inserting Eq.~\ref{eq:gen_gauge_nondeg} into Eq.~\ref{eq:numer_scheid} trivially demonstrates its degenerate band gauge invariance.

Next, we compare our Eq.~\ref{eq:ife_all_three} with the expression for IFE from Refs.~\onlinecite{Battiato2014} and \onlinecite{Berritta2016}. The analogous doubly resonant time-independent contribution to the IFE from Refs.~\onlinecite{Battiato2014} and \onlinecite{Berritta2016} can be rewritten in the form of our Eq.~\ref{eq:ife_simpler} with the following choice of the numerator $g_{nm}^{\rm elec}$,
\begin{align}\label{eq:numer_berritta}
\sum_{N=1}^2
\sum_{M=1}^2
\braket{\phi_{nN} | V | \phi_{mM}} 
\braket{\phi_{mM} | M^{\rm spin} | \phi_{mM}}
\braket{\phi_{mM} | V^{\dagger} | \phi_{nN}}.
\end{align}
Eq.~\ref{eq:numer_berritta} contains only two sums over the degenerate indices ($N$ and $M$), while our Eq.~\ref{eq:numer} has a sum over three degenerate indices ($N$, $M$, and $M'$). However, since matrix elements of $V$ and $M^{\rm spin}$ still transform as matrices under degenerate band gauge transformation given in Eq.~\ref{eq:gen_gauge}, the Eq.~\ref{eq:numer_berritta} is therefore not in the form of a trace of a product of matrices, and therefore Eq.~\ref{eq:numer_berritta} is not degenerate band gauge-invariant. This is why in Ref.~\onlinecite{Berritta2016}, authors had to work in a special gauge to obtain a gauge-invariant IFE.\footnote{Private communication with P.~M.~Oppeneer.}
In this special gauge, the operator $M^{\rm spin}$ must be chosen to be diagonal in the space of each doubly degenerate band. Such an approach, in a somewhat different context, of using a gauge in which $M^{\rm spin}$ is diagonal, has also been proposed in Ref.~\onlinecite{Pientka2012}. 

Clearly, the numerators given in Eqs.~\ref{eq:numer} and \ref{eq:numer_berritta} are identical if $\braket{\phi_{mM} | M^{\rm spin} | \phi_{mM'}}$ is diagonal in the doublet indices $M$ and $M'$.  However, first-principles codes do not necessarily provide an output electron wavefunction in a gauge where $M^{\rm spin}$ is diagonal. Therefore, the use of a degenerate band gauge-invariant expression such as Eq.~\ref{eq:numer} is preferred, as it gives the same result regardless of the gauge choice. 

Furthermore, the eight contributions to the non-doubly-resonant $M^{\rm IFE}_{\rm ndr}$, given in Appendix~\ref{app:nonresonant}, contain matrix elements of $M^{\rm spin}$ between electronic states with different energies. However, the gauge freedom of the form given in Eq.~\ref{eq:gen_gauge} does not allow a mixture of states with different energies. Therefore, it is generally not possible to find a special degenerate band gauge to evaluate $M^{\rm IFE}_{\rm ndr}$, as in the case of $M^{\rm IFE}_{\rm elec}$ and $M^{\rm IFE}_{\rm hole}$. Instead, one must evaluate $M^{\rm IFE}_{\rm ndr}$ using the manifestly degenerate band gauge invariant form, as given in Appendix~\ref{app:nonresonant}. 

\subsection{\label{sec:numerical_test}Numerical test of degenerate band gauge invariance}

\begin{figure}[!t]
    \centering
    \includegraphics{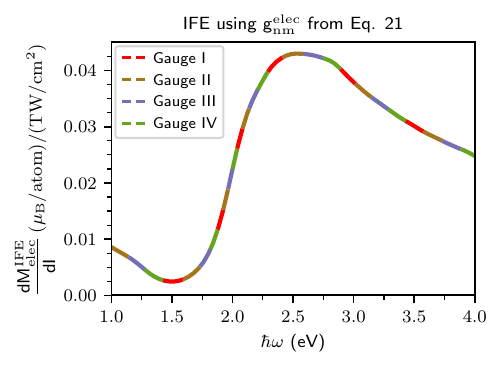}
    \caption{\label{fig:deg2} $M^{\rm IFE}_{\rm elec}$ of Au as a function of incoming light frequency $\omega$ computed using Eq.~\ref{eq:numer} and Eq.~\ref{eq:ife_simpler} (or, equivalently, Eq.~\ref{eq:ife_elec}) in four different degenerate band gauges. All four degenerate band gauge choices give the same numerical value of $M^{\rm IFE}_{\rm elec}$.}
\end{figure}

\begin{figure}[!t]
    \centering
    \includegraphics{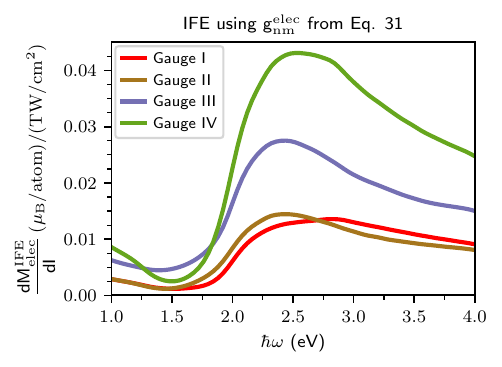}
    \caption{\label{fig:deg1} Same as Fig.~\ref{fig:deg2} but using $g_{nm}^{\rm elec}$ from Eq.~\ref{eq:numer_berritta} instead of Eq.~\ref{eq:numer}. Now four different degenerate band gauge choices give four different values of $M^{\rm IFE}_{\rm elec}$.  The green curve in this figure, corresponding to gauge choice IV, is numerically identical to the green curve in Fig.~\ref{fig:deg2}.}
\end{figure}

We now numerically test the degenerate band gauge invariance of various formulations of IFE. For this test, we consider the bulk fcc gold, as it is an inversion-symmetric non-magnetic material, so it has $\cal{PT}$ symmetry. We perform density functional theory calculations using Quantum ESPRESSO.\cite{Giannozzi2017}  We use the generalized gradient approximation\cite{Perdew_1996} to density functional theory. Atomic potentials are replaced with the fully-relativistic ONCV pseudopotentials~\cite{Hamann2013} from the pseudo-dojo library.\cite{VANSETTEN201839} We set the kinetic energy cutoff for the planewave basis expansion to 120~Ry. We perform self-consistent calculations using a $k$-mesh grid of $28\times28\times28$ and non-self-consistent calculations on a $8\times8\times8$ grid. We used an experimental lattice constant of 4.08~\AA{}.\cite{Patel_1967}  

We employ the maximally localized Wannier functions approach~\cite{Marzari2012} as implemented in Wannier90\cite{Pizzi2020} to compute the IFE by Wannier interpolation.\cite{Wang2007} We use atom-centered orbitals $sp^3d^2$, $d_{xy}$, $d_{yz}$, and $d_{zx}$ to construct Wannier functions. For the Wannier interpolation, we used a very flexible interface of the Wannier~Berri\cite{Stepan2021} package that enabled us to implement the calculation of the IFE.  We find that a $k$-point interpolation grid of $100\times100\times100$ is sufficient to give the converged result. To avoid singular points in the Brillouin zone with high symmetry, we shifted the uniform interpolation grid by a small random displacement along all three Cartesian directions.  In most of our calculations, we use a constant inverse lifetime of $\eta=0.1$~eV.

To check the degenerate band gauge invariance of Eqs.~\ref{eq:ife_elec} and \ref{eq:numer_berritta}, we first discuss different strategies we used to obtain distinct gauge choices for the electron orbitals. Within the Kohn-Sham formalism,~\cite{Kohn1965} electron orbitals are eigenvectors of the differential equation that has the form of a Schrodinger-like Eq.~\ref{eq:unpert_ham}. By the periodicity of the solid, we need to solve one Schrodinger-like equation for each $\bm k$-point in the Brillouin zone. To obtain different degenerate band gauge choices for the solutions of this equation, we will use to our advantage the fact that our numerical solutions to Eq.~\ref{eq:unpert_ham} are based on an iterative diagonalization procedure. These iterative diagonalization procedures start from a user-provided initial guess for the electron wavefunction. The initial guess is then iteratively optimized by the diagonalization algorithm. Clearly, different choices of the initial guess for the wavefunction will then be iterated by the diagonalization procedure to a different choice of the final wavefunctions. When solutions to Eq.~\ref{eq:unpert_ham} at a single $\bm k$-point are non-degenerate, the only potential difference in the final resulting wavefunctions is the overall phase factors. These phase factors have the form of the trivial gauge transformation as in Eq.~\ref{eq:gen_gauge_nondeg} and do not pose any difficulties in calculating the IFE. 

On the other hand, if Eq.~\ref{eq:unpert_ham} at a single $\bm k$-point has degenerate solutions, then the just described iterative diagonalization procedure will yield an arbitrary linear combination of degenerate electron orbitals. In the case of $\cal{PT}$ symmetric material, where the bands are at least two-fold degenerate at each $\bm k$, these linear combinations of degenerate electron orbitals result in gauge freedom described by a $2 \times 2$ unitary matrix. This is precisely the gauge transformation Eq.~\ref{eq:gen_gauge} we discussed in the previous subsection. We note that the electron density is clearly unchanged by the gauge choice described above, so any physical property, such as total energy, that depends on the electron density is gauge invariant.

In the following, we discuss four strategies that we use to obtain four different gauge choices. We label these gauges as I, II, III, and IV.

We obtain gauge-I by initializing the electron orbital to a linear combination of atomic-like orbitals centered on Au atoms.  On the other hand, we obtain gauge-II by initializing the electron orbital to a completely random linear combination of plane-wave basis functions. In particular, both spinor components of the electron orbital are randomized, which results in the random initial orientation of spinors.
 
For constructing gauges III and IV, we use a somewhat different approach. Instead of specifying different initial guesses for the electron orbital, we apply a weak external perturbation to the system, which then effectively nudges the iterative procedure towards a specific choice of the gauge. For constructing gauge-III, we chose as our external perturbation a Zeeman field pointing along the $x$-axis. Such a Zeeman field will induce a small splitting of the $\cal{PT}$-degenerate electronic bands on the order of 10$^{-6}$~eV. Therefore, regardless of the initial choice of the electron orbital, the final electron orbital will always correspond (up to an overall phase factor) to the orbitals with spinors aligned along the $x$-axis. However, for the purposes of our test of the gauge invariance of IFE, we still must treat the bands split by 10$^{-6}$~eV as if they were degenerate doublets. We confirmed that these small energy splittings do not affect the computed value of the IFE by varying the strength of the applied Zeeman field. In particular, we find that the IFE is nearly unchanged even if we use a stronger Zeeman field that induces 10 or 100 times larger splitting of the degenerate doublets (so that the splittings are on the order of 10$^{-5}$~eV or 10$^{-4}$~eV). Finally, we construct gauge-IV similar to gauge-III, but we apply the small Zeeman field along the $z$-axis instead of the $x$-axis.

Figure~\ref{fig:deg2} shows the IFE that we calculated using our Eq.~\ref{eq:ife_elec}. The horizontal axis denotes the frequency of the incoming light, $\omega$. The different colored lines in the figure correspond to the IFE calculated using gauges I, II, III, and IV. As can be seen from the figure, all four gauges result in a nearly identical value of IFE. The relative numerical differences between the IFE calculated in these four cases are at most $10^{-4}$. This numerical test further confirms the gauge invariance of Eq.~\ref{eq:ife_elec} that we demonstrated analytically in the previous section. We also numerically verified the degenerate band gauge invariance of $M^{\rm IFE}_{\rm hole}$ and each of the eight terms that contribute to $M^{\rm IFE}_{\rm ndr}$.

We remind the reader that Eq.~\ref{eq:ife_elec} can be written equivalently in the form of Eq.~\ref{eq:ife_simpler} with the numerator $g_{nm}^{\rm elec}$ taken from Eq.~\ref{eq:numer}. This form is useful in comparing our equation with that from previous work. In particular, a doubly-resonant time-independent contribution to IFE from Refs.~\onlinecite{Battiato2014} and \onlinecite{Berritta2016} can be written as Eq.~\ref{eq:ife_simpler} but with numerator $g_{nm}^{\rm elec}$ from Eq.~\ref{eq:numer_berritta} instead of Eq.~\ref{eq:numer}. (At the moment, we neglect a different functional form of the denominator, as it does not affect the gauge invariance. We discuss the role of different denominators in Sec.~\ref{sec:num_comparison}.) Figure~\ref{fig:deg1} shows the IFE that we calculated using the numerator from Eq.~\ref{eq:numer_berritta} and the denominator from our Eq.~\ref{eq:numer}. As can be seen from the figure, four different degenerate band gauge choices result in four different values of IFE, which is non-physical. This numerical test clearly demonstrates the dependence of Eq.~\ref{eq:numer_berritta} on the choice of degenerate band gauge. However, as discussed in Sec.~\ref{sec:previous}, numerators from Eq.~\ref{eq:numer} and Eq.~\ref{eq:numer_berritta} are equivalent if one chooses a degenerate band gauge in which the matrix elements of $M^{\rm spin}$ are diagonal in the doublet indices. To numerically test this equivalence, we consider our results in the case of gauge IV, as in this gauge $M^{\rm spin}$ is forced to be diagonal by an application of a small Zeeman field along the $z$-axis. Indeed, as can be seen by comparing the green line in Figs.~\ref{fig:deg2} and \ref{fig:deg1}, the gauge-IV indeed leads to the same doubly-resonant $M^{\rm IFE}_{\rm elec}$ using Eq.~\ref{eq:numer} or Eq.~\ref{eq:numer_berritta}. Unfortunately, as discussed at the end of Sec.~\ref{sec:previous}, it is not possible to use a special gauge (such as gauge IV) to calculate the non-doubly resonant $M^{\rm IFE}_{\rm ndr}$.

\subsection{\label{sec:non-doubly-res} Non-doubly resonant contributions to IFE}

\begin{figure}[!t]
    \centering
    \includegraphics{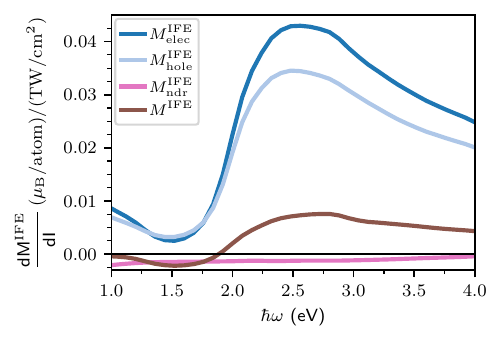}
    \caption{\label{fig:four-terms} Decomposition of the interband spin IFE of Au $M^{\rm IFE}$ (brown color) into doubly-resonant contributions $M^{\rm IFE}_{\rm elec}$ (dark blue, Eq.~\ref{eq:ife_elec}) and $\left|M^{\rm IFE}_{\rm hole}\right|$ (light blue, Eq.~\ref{eq:ife_hole}), as well as the non-doubly resonant contribution $M^{\rm IFE}_{\rm ndr}$ (pink color, Eq.~\ref{eq:ndr}). Near resonance, the non-doubly resonant terms are negligible compared to the doubly-resonant terms.  We use $\eta=0.1$~eV.}
\end{figure}

In Eq.~\ref{eq:ife_all_three} we decomposed total IFE into $M^{\rm IFE}_{\rm elec}-M^{\rm IFE}_{\rm hole}$ which can be doubly resonant, as well as $M^{\rm IFE}_{\rm ndr}$ which can't. Therefore, in a material, such as bulk gold, in which there is a clear resonance feature in the band structure in the optical regime, we expect that the non-doubly-resonant term $M^{\rm IFE}_{\rm ndr}$ will be small in magnitude compared to $M^{\rm IFE}_{\rm elec}-M^{\rm IFE}_{\rm hole}$. In fact, this is what we find numerically for the interband non-doubly-resonant terms, as shown in Fig.~\ref{fig:four-terms}. 

Now we discuss various non-doubly-resonant contributions to IFE. The first group of non-doubly-resonant contributions to IFE are eight interband contributions to $M^{\rm IFE}_{\rm ndr}$. These are listed in the Appendix~\ref{app:nonresonant}. Figure~\ref{fig:non-and-resonant} shows that, as expected, in the resonance region (around 2~eV) these contributions to the IFE are about 100~times smaller than the doubly-resonant contributions. At lower frequencies, around 1~eV the non-doubly-resonant terms are only about 5 times smaller.

\begin{figure}[!t]
    \centering
    \includegraphics{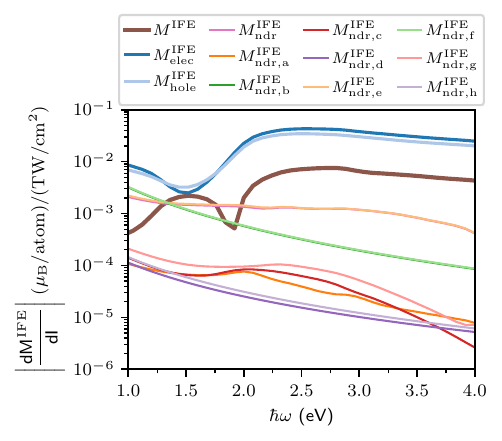}
    \caption{\label{fig:non-and-resonant} All doubly-resonant (Eqs.~\ref{eq:ife_elec} and \ref{eq:ife_hole}) and non-doubly-resonant (Eq.~\ref{eq:ndr}) interband spin components of IFE in bulk gold.  Vertical scale is logarithmic. We use $\eta=0.1$~eV.}
\end{figure}

The second group of contributions are those that involve intraband transitions. However, in a nonmagnetic inversion-symmetric material these intraband transitions are never doubly resonant. To show this, let us consider for a moment the doubly resonant interband contribution $M^{\rm IFE}_{\rm elec}-M^{\rm IFE}_{\rm hole}$ given in Eqs.~\ref{eq:ife_elec} and \ref{eq:ife_hole}. To obtain the intraband contribution, we need to consider $n$ and $m$ corresponding to the same band, and we should compute the matrix element of $V$ using Eq.~\ref{eq:intraband}. However, since Eq.~\ref{eq:intraband} is diagonal in the doublet indices, it is easy to show that the resulting doubly resonant intraband contribution to the IFE is proportional to the trace of $M^{\rm spin}$ over the doublet. In a nonmagnetic inversion-symmetric material, such as bulk Au, this trace is zero. On the other hand, if one considers the remaining eight non-doubly-resonant interband contributions to the IFE, given in Appendix~\ref{app:nonresonant}, one can easily see that these do not vanish. However, since these are non-doubly resonant, we expect them to have a negligible contribution to the total IFE in the optical range of $\omega$.

The third group of non-doubly resonant contributions to the IFE are those that result in a time-dependent magnetization that oscillates with twice the frequency ($2\omega$) of the incoming light. However, as can be seen from Eq.~\ref{eq:B2}, none of these contributions can be made doubly resonant, since the energy denominators can generically never be both zero for any set of bands. Therefore, the magnitude of the magnetization oscillating with frequency $2\omega$ is going to be negligible to the time-independent contribution to IFE. The same conclusion was also made in Refs.~\onlinecite{Popova2011,Battiato2014}.

\section{\label{sec:discussion}Discussion}

Now we discuss in more detail our results for the spin-contribution of IFE in Au.

\subsection{\label{sec:intuit} Intuitive picture of the doubly resonant IFE}

In Eq.~\ref{eq:ife_all_three} we decomposed IFE into a doubly resonant part ($M^{\rm IFE}_{\rm elec} - M^{\rm IFE}_{\rm hole}$) and the non-doubly resonant part ($M^{\rm IFE}_{\rm ndr}$). Now we are going to rewrite the dominant, doubly-resonant, contribution in a way that more clearly demonstrates the meaning of $M^{\rm IFE}_{\rm elec}$ and $-M^{\rm IFE}_{\rm hole}$. First, we adopt a degenerate band gauge in which the matrix element $\braket{\phi_{mM} | M^{\rm spin} | \phi_{mM'}}$ is diagonal in $M$ and $M'$ for each $m$. We will denote states in such a basis as $\ket{\widetilde{\phi}_{mM}}$. It is then easy to show that $M^{\rm IFE}$ can be written as,
\begin{widetext}
\begin{align}
M^{\rm IFE}_{\rm elec} 
-
M^{\rm IFE}_{\rm hole} 
= 
\int_{\rm BZ} 
\frac{d^3k}{(2\pi)^3} 
\sum_n^{\rm occ} 
\sum_{N=1}^2
\sum_m^{\rm emp}
\sum_{M=1}^2
\frac{
\braket{\widetilde{\phi}_{nN} | V | \widetilde{\phi}_{mM}} 
\left[
\braket{\widetilde{\phi}_{mM} | M^{\rm spin} | \widetilde{\phi}_{mM}}
-
\braket{\widetilde{\phi}_{nN} | M^{\rm spin} | \widetilde{\phi}_{nN}}
\right]
\braket{\widetilde{\phi}_{mM} | V^{\dagger} | \widetilde{\phi}_{nN}} }
{(E_m - E_n - \hbar \omega)^2 + \eta^2}.
\label{eq:rewritten_IFE}
\end{align}
\end{widetext}
Therefore, the doubly resonant contribution to $M^{\rm IFE}$ can be interpreted as a process of optical excitation and de-excitation of a solid with circularly polarized light. These optical processes are weighted by the effective magnetic moment of excitations. According to Eq.~\ref{eq:rewritten_IFE}, the effective magnetic moment of excitation consists of the magnetic moment of the excited electron $\braket{\widetilde{\phi}_{mM} | M^{\rm spin} | \widetilde{\phi}_{mM}}$ and the excited hole $-\braket{\widetilde{\phi}_{nN} | M^{\rm spin} | \widetilde{\phi}_{nN}}$. Clearly, to have a large IFE one needs to find a material in which, in addition to the resonance, there is as large an asymmetry as possible between the effective magnetic moment of the electron and the hole.

\subsection{Comparison of IFE to the optical spectrum}

As shown in Fig.~\ref{fig:four-terms}, we find that the total spin IFE has a resonance-like peak when $\omega$ is around 2.5~eV. We assign this peak to the interband transitions from $d$ to $sp$-like states. This dependence of IFE on frequency is reminiscent of the well-known similar frequency dependence of the optical dielectric function.\cite{Theye_1970,Christensen_1971} Clearly, at some fixed $\omega$, both our Eqs.~\ref{eq:ife_elec} and \ref{eq:ife_hole}, as well as the dielectric function increase if there is a pair of states separated by $\hbar \omega$. In fact, the dielectric function shows a similar dependence on $\omega$ as the joint density of the states (JDOS) divided by $\omega^2$, as discussed, for example, in Refs.~\onlinecite{Theye_1970,Christensen_1971,Bordoloi_1988}. Therefore, it is tempting to compare our calculated IFE value with JDOS$/\omega^2$. Nevertheless, as discussed in the previous subsection, the dominant spin contribution to IFE in gold is a result of compensating magnetic moments of excited electrons and holes. Therefore, we first focus on the comparison of JDOS with $M^{\rm IFE}_{\rm elec}$ and then with $M^{\rm IFE}_{\rm hole}$.

\begin{figure}[!t]
    \centering
    \includegraphics{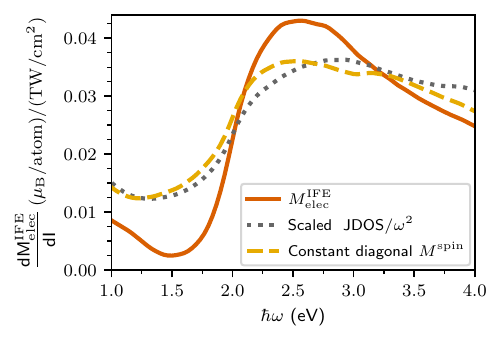}
    \caption{\label{fig:au_m_const_jdos} Comparison of $M^{\rm IFE}_{\rm elec}$ for Au calculated using Eq.~\ref{eq:ife_elec} (solid red line) with two approximants: scaled JDOS/$\omega^2$ (dotted gray line) and $M^{\rm IFE}_{\rm elec}$ computed with assuming constant diagonal matrix elements of $M^{\rm spin}$ from Eq.~\ref{eq:approxM_elec} (dashed yellow line). Here we use ${\cal M}_{\rm elec}$ equal to 0.22~$\mu_B$ per one Au atom per TW/cm$^2$.}
\end{figure}

We define JDOS as,
\begin{align}\label{eq:jdos}
    \text{JDOS} (\hbar \omega) = & 4 \int_{BZ} \frac{d^3k}{(2\pi)^3} \sum_n^{\rm occ} \sum_m^{\rm emp} \delta \left(E_{m\bm{k}} - E_{n\bm{k}} - \hbar \omega\right).
\end{align}
In the numerical calculation, we replaced the Dirac delta function with a Lorentzian with a width of 0.1~eV. Figure~\ref{fig:au_m_const_jdos} compares the $M^{\rm IFE}_{\rm elec}$ part of IFE (solid red line) with JDOS/$\omega^2$ (dotted gray line). We rescaled JDOS /$\omega^2$ with an arbitrary constant prefactor to make it visually easier to compare the curves. The JDOS/$\omega^2$ shows a resonance-like structure near 2.5~eV, just like our calculated $M^{\rm IFE}_{\rm elec}$.
Clearly, the spin component of IFE in a simple metal like bulk gold mostly comes from the band-structure effects. Furthermore, it seems likely that the matrix element of $M^{\rm spin}$ appearing in Eq.~\ref{eq:ife_elec}, but not in JDOS, weakly depends on the $\bm k$-point in bulk Au. To test this hypothesis, we performed a somewhat simplified IFE calculation in which we assume that the $M^{\rm spin}$ matrix is proportional to a $2\times2$ identity matrix in the doublet indices,
\begin{align}
& \braket{\phi_{mM {\bm k}} | M^{\rm spin} | \phi_{mM' {\bm k}}} \longrightarrow {\cal M}_{\rm elec} \delta_{M M'},
\label{eq:approxM_elec}
\end{align}
Here, a single numerical constant ${\cal M}_{\rm elec}$ is the same for all empty states labeled by $m$, $M$, $M'$, and $\bm{k}$. We note that in this somewhat simplified calculation, we are replacing a traceless matrix $\braket{\phi_{mM {\bm k}} | M^{\rm spin} | \phi_{mM' {\bm k}}}$ with the identity matrix with a non-zero trace. Therefore, effectively, in this simplified approach, we treat gold as if it had spin-polarized electronic states and the spin magnetic moment of each empty electronic state is ${\cal M}_{\rm elec}$. However, we find this replacement to be convenient as it is gauge invariant and can be parameterized with a single number, ${\cal M}_{\rm elec}$. With replacement from Eq.~\ref{eq:approxM_elec}, we find that
$$
{\cal M}_{\rm elec} = 0.22 \ (\mu_{\rm B} / {\rm atom}) / ({\rm TW} / {\rm cm}^2 )
$$ 
gives the final result that is numerically as similar as possible to the $M^{\rm IFE}_{\rm elec}$ computed using actual matrix elements of $M^{\rm spin}$ (compare yellow dashed line and solid red line in Fig.~\ref{fig:au_m_const_jdos}). Given this good agreement, we can now consider ${\cal M}_{\rm elec}$ as an intrinsic measure of the average effective magnetic moment of the electron in Au.

However, as just discussed, the total spin IFE in Au consists not only of $M^{\rm IFE}_{\rm elec}$ but also of $-M^{\rm IFE}_{\rm hole}$ and $M^{\rm IFE}_{\rm ndr}$.  The magnitude of $M^{\rm IFE}_{\rm ndr}$ is negligible compared to $M^{\rm IFE}_{\rm elec}$ so we can ignore it. However, $M^{\rm IFE}_{\rm hole}$ is similar in magnitude to $M^{\rm IFE}_{\rm elec}$. In fact, we can once again reproduce most spectral features of $-M^{\rm IFE}_{\rm hole}$ if we replace the matrix element of $M^{\rm spin}$ with another constant,
\begin{align}
& \braket{\phi_{nN {\bm k}} | M^{\rm spin} | \phi_{nN' {\bm k}}} \longrightarrow {\cal M}_{\rm hole} \delta_{N N'}.
\label{eq:approxM_hole}
\end{align}
If we determine ${\cal M}_{\rm hole}$ analogously to ${\cal M}_{\rm elec}$ we find that the magnetic moment of the hole is somewhat smaller in magnitude, 
$$
{\cal M}_{\rm hole} = 0.18 \ (\mu_{\rm B} / {\rm atom}) / ({\rm TW} / {\rm cm}^2).
$$
Therefore, as discussed in Sec.~\ref{sec:intuit}, the non-zero IFE is a result of incomplete cancellation between the effective magnetic moment of the electron and the hole ($0.22$ versus $0.18 \ (\mu_{\rm B} / {\rm atom}) / ({\rm TW} / {\rm cm}^2$)).

\subsection{Role of electron lifetime}

\begin{figure}[!t]
    \centering
    \includegraphics{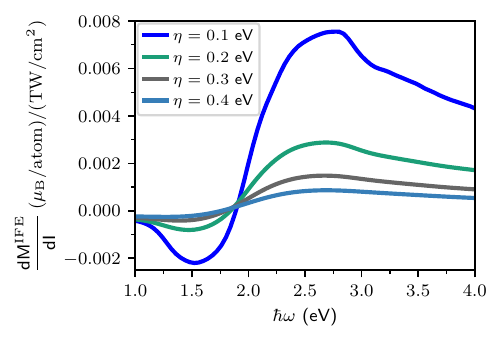}
    \caption{\label{fig:au_eta_vary}Comparison of total spin IFE for Au with different choice of $\eta$.}
\end{figure}

We now briefly comment on the importance of the electron lifetime ($\eta$) magnitude used in our calculation. Our results shown so far were obtained with $\eta=0.1$~eV. In Fig.~\ref{fig:au_eta_vary}, we show the calculated value of the total spin IFE using $\eta$ set to 0.2, 0.3, and 0.4 eV. Comparing the IFE at these lifetimes, we observe that the basic features of the IFE remain the same, but the overall magnitude is reduced near the resonance. This is consistent with the expectation that a larger $\eta$ would lead to a less pronounced resonance structure of the IFE.

\subsection{Comparison with numerical values from the previous work}
\label{sec:num_comparison}

Now we compare the numerical values of the spin component of IFE in bulk gold calculated in this work with those reported in Ref.~\onlinecite{Berritta2016}. In our work, we find that the spin part of IFE in gold closely resembles the optical properties of gold. That is, IFE in this case is large when $\omega$ is close to the $d$--$s$ optical resonance. On the other hand, the spin component of IFE reported in Ref.~\onlinecite{Berritta2016} does not show this resonant feature. Instead, the IFE in that work can be reasonably well approximated as $\omega^{-1}$ in the range of energies from 1.0 to 4.0~eV. We now discuss the likely origin of the difference in the computed value of IFE. The most significant difference comes from the subtle difference in the relative signs of $i \eta^{-1}$ in the energy denominators. As can be seen in Eqs.~\ref{eq:ife_elec} and \ref{eq:ife_hole}, the parameter $\eta$ introduces a Lorentzian-like broadening of the optical transitions in our work, so that
\begin{align}
M^{\rm IFE}_{\rm elec} - M^{\rm IFE}_{\rm hole} \sim \sum_{\ldots} \frac{\ldots}{ (E_m - E_n - \hbar \omega)^2 + \eta^2}.
\end{align}
The same line shape is used in Ref.~\onlinecite{Scheid2019} where it was derived from the application of the Fermi golden rule.  More specifically, the application of the Fermi golden rule from Ref.~\onlinecite{Scheid2019} results in an IFE that is proportional to the Dirac delta function,
\begin{align}
 \sim \lim_{\eta \rightarrow 0} \sum_{\ldots}\frac{\ldots}{ (E_m - E_n - \hbar \omega)^2 + \eta^2}.
\end{align}

In our work, the Lorentzian-like line shape follows from the opposite sign of $i\eta$ in the denominators of $B^{(1)}_j$ and its complex conjugate $\left[B^{(1)}_j\right]^{*}$. This is consistent with Refs.~\onlinecite{Ward1965}, \onlinecite{Buckingham2000}, and \onlinecite{Norman2006} as discussed in Appendix~\ref{app:perturbation}.

In Refs.~\onlinecite{Battiato2014} and \onlinecite{Berritta2016}, the functional form of the energy denominator results in a non-Lorentzian line shape which removes resonant features of IFE near the $d$--$s$ optical resonance. In particular, the spectral form in the quantum-mechanical derivation in Refs.~\onlinecite{Battiato2014} and \onlinecite{Berritta2016} is of the following form,
\begin{align}
 \sim {\rm Re} \sum_{\ldots} \left[ \frac{\ldots}{ (E_m - E_n - \hbar \omega + i \eta)^2} \right].
\end{align}
This function has a minimum at $ \hbar \omega = E_m - E_n$ and two maximums at $ \hbar \omega = E_m - E_n \pm \sqrt{3} \eta$. However, the integral of this function over $\omega$ is zero.  Therefore, the convolution of this spectral form with the resonance in the gold band structure will generically wash out the resonance in IFE. In fact, for a sufficiently small $\eta$, smaller than any feature in the band structure, the total computed IFE will tend to zero. We note that, on the other hand, the semi-classical derivation from the same work (Eq.~10 in Ref.~\onlinecite{Battiato2014}) finds a Lorentz-like behavior of the IFE, in agreement with our work, as well as with Ref.~\onlinecite{Scheid2019}.

\section{\label{sec:summary}Summary}

In this work, we revisit the IFE theory for non-magnetic metals having inversion symmetry. In such material, the electronic bands are at least two-fold degenerate everywhere in the Brillouin zone. We show that our expression for IFE is degenerate band gauge invariant in the subspace formed by two-fold degenerate bands. We demonstrated the degenerate band gauge invariance of our approach both analytically and numerically. More generally, a similar concern with the degeneracy of electronic bands is going to be relevant for other spin-orbit driven properties of the material that require one to go beyond the first order in perturbation theory. As we discussed in Sec.~\ref{sec:derivation}, the leading term for IFE is of the order $\lambda^2$. Therefore, the final expression for IFE includes a sum over the triplet of electronic states.  The difficulty with degeneracy then occurs whenever a pair of these states are both occupied or both empty, as they might then correspond to the same doubly degenerate manifold.  We note that physical properties, such as the spin-Hall effect, which also occurs in materials with two-fold degenerate bands, are of the order $\lambda^1$. Therefore, in computing the spin-Hall effect, for example, one does not need to be as careful in dealing with the degenerate band gauge invariance within the space of doubly degenerate bands. In other words, the expression for the spin-Hall effect has only two sums over the electronic states, one of which is empty and another is occupied, so these states must correspond to distinct doubly degenerate manifolds.

When analyzing the dependence of the IFE on the frequency $\omega$ of the incoming light, we find that the spin component of our calculated IFE in gold shows a resonance-like structure around 2.5~eV in contrast to the findings of Ref.~\onlinecite{Berritta2016}. We assign the resonance structure to the $d$--$s$ electronic transitions, which are also present in the dielectric function of gold. The situation in other transition metals is somewhat more involved than in the case of gold, and this will be a topic of future work.

Finally, while the current work focuses on the spin component of the IFE, we suspect that for a more direct quantitative comparison with the experiment, it is important to include the orbital component of the IFE as well. In particular, we expect that the intraband contribution to the orbital part of IFE might be large in simple metals such as bulk gold, as suggested by the semi-classical theories of IFE,\cite{HERTEL2006L1, Hurst2018} as well as by a quantum-mechanical calculation of the local part of the orbital moment.\cite{Battiato2014,Berritta2016} However, as discussed earlier, unfortunately, the intraband optical part of IFE is the one that is hardest to formulate on a sound quantum-mechanical footing for an infinite bulk periodic solid.

\begin{acknowledgements}
This work benefited from the development of a very flexible computer package Wannier~Berri by S.~Tsirkin.\cite{Stepan2021} This work was supported by the U.S. Army Research Office under Grant No. W911NF-20-1-0274. The authors acknowledge discussions with R.~Wilson, L.~Vuong, S.~Tsirkin, and I.~Souza.
\end{acknowledgements}

\appendix

\section{\label{app:light-electron} Electron-light interaction term}

The electric field $\bm E$ of circularly polarized light propagating along the $\hat{\bm z}$ axis, with wavevector $q$, frequency $\omega$, and intensity $I$ is given as,
\begin{align}
{\bm E} (x,y,z,t) = \sqrt{\frac{I}{\epsilon_0 c}} \left[ \hat{\bm x} \cos \left( q z - \omega t \right) + \hat{\bm y} \sin \left( q z - \omega t \right)  \right].
\end{align}
The interaction of this electric field with the electron in the solid can be introduced following various strategies, as discussed, for example, in Refs.~\onlinecite{Blount1962,Sipe1995,Ventura2017,CALDERIN2017,Bouldi2017,Milonni1989}. However, all of these approaches lead to almost the same numerical result, as long as $\hbar \omega$ is larger than our effective broadening $\eta$.\cite{Mermin1970} 

The interaction of electrons with the electric field $\bm E$ is described with the following form of the perturbation $V$ appearing in Eq.~\ref{eq:full_ham},
\begin{align}
V = - i \frac{e}{2 \omega} \sqrt{\frac{I}{\epsilon_0 c}} \left( v_x + i v_y \right).
\end{align}
Here $v_x$ and $v_y$ are velocity operators. For the interband matrix elements of $v_x$ (and similarly $v_y$) Ref.~\onlinecite{Blount1962} gives,
\begin{align}
\braket{\phi_{nN} | v_x | \phi_{mM}} 
\longrightarrow
\frac{i}{\hbar}
(E_n - E_m)
A_{nNmM}^x
\textrm{\quad (interband)}.
\end{align}
The matrix elements of $V$ are then easily computed, as shown in Eq.~\ref{eq:Voper} in the main text. The equivalent contribution for the intraband transition is,
\begin{align}
\braket{\phi_{nN} | v_x | \phi_{nN'}} 
\longrightarrow
\frac{1}{\hbar}
\frac{\partial E_n}{\partial k_x}
\delta_{N N'}.
\end{align}
which results in the following intraband matrix element of $V$,
\begin{align}
\label{eq:intraband}
& \braket{\phi_{nN} | V | \phi_{nN'}} 
= - i \frac{e}{2} \sqrt{\frac{I}{\epsilon_0 c}} 
\frac{1}{\hbar \omega}
\left( 
\frac{\partial E_n}{\partial k_x}
 + i 
\frac{\partial E_n}{\partial k_y}
 \right) \delta_{N N'}. 
\end{align}

\begin{widetext}
\section{\label{app:perturbation} Perturbative expansion}

Following Refs.~\onlinecite{Langhoff1972, Bhattacharyya1986}, the lowest-order
perturbative expansion for $\alpha_{ji}$ is given as
\begin{align}
\hbar\alpha_{ji} & = (E_j - E_i) + \lambda^2
\left[\sum_{l}^{l \neq i} \left(\frac{\braket{ \Phi_l | V | \Phi_i}\braket{ \Phi_i | V^\dagger | \Phi_l}}{\hbar \alpha_{li} + \hbar \omega} 
+\frac{\braket{ \Phi_i | V | \Phi_l} \braket{ \Phi_l | V^\dagger | \Phi_i}}{\hbar \alpha_{li} - \hbar \omega} \right)-\frac{\braket{ \Phi_i | V | \Phi_i}\braket{ \Phi_i | V^\dagger | \Phi_i}}{\hbar \alpha_{ji}} \right].
\label{eq:alpha_nm}
\end{align}
while the expansion of $B_j$ is
\begin{align}
 B^{(1)}_j & = - \left( \frac{\braket{ \Phi_j | V | \Phi_i}}{\hbar \alpha_{ji} + \hbar \omega + i \eta} e^{i \omega t } + \frac{\braket{ \Phi_j | V^\dagger | \Phi_i}}{\hbar \alpha_{ji} - \hbar \omega + i \eta} e^{- i \omega t}\right) e^{i \alpha_{ji} t}
 \label{eq:B1}
\\
B^{(2)}_j & =  \left\{ \left[\sum_{l}^{l \neq i} \frac{\braket{ \Phi_j | V | \Phi_l}\braket{ \Phi_l | V | \Phi_i}}{\hbar \alpha_{li} + \hbar \omega + i \eta} -\frac{\braket{ \Phi_i | V | \Phi_i}\braket{ \Phi_j | V | \Phi_i}}{\hbar \alpha_{ji} + \hbar \omega + i \eta} \right] \frac{e^{2 i \omega t }} {\hbar \alpha_{ji} + 2 \hbar \omega + 2 i \eta} \right.  \nonumber \\
 & + \left[\sum_{l}^{l \neq i}
 \left(
 \frac{\braket{ \Phi_j | V | \Phi_l}\braket{ \Phi_l | V^\dagger | \Phi_i}}{\hbar \alpha_{li} - \hbar \omega + i \eta} 
+\frac{\braket{ \Phi_j | V^\dagger | \Phi_l}\braket{ \Phi_l | V | \Phi_i}}{\hbar \alpha_{li} + \hbar \omega + i \eta} 
\right)
-\frac{\braket{ \Phi_i | V | \Phi_i}\braket{ \Phi_j | V^\dagger | \Phi_i}}{\hbar \alpha_{ji} - \hbar \omega + i \eta} -\frac{\braket{ \Phi_i | V^\dagger | \Phi_i}\braket{ \Phi_j | V | \Phi_i}}{\hbar \alpha_{ji} + \hbar \omega + i \eta}\!  \right]\! \!  \frac{1} {\hbar \alpha_{ji} + 2 i \eta} \nonumber \\ 
&+ \left. \left[ \sum_{l}^{l \neq i} \frac{\braket{ \Phi_j | V^\dagger | \Phi_l}\braket{ \Phi_l | V^\dagger | \Phi_i}}{\hbar \alpha_{li} - \hbar \omega + i \eta} -\frac{\braket{ \Phi_i | V^\dagger | \Phi_i}\braket{ \Phi_j | V^\dagger | \Phi_i}}{\hbar \alpha_{ji} - \hbar \omega + i \eta} \right] \frac{e^{- 2 i \omega t }}{\hbar \alpha_{ji} - 2 \hbar \omega + 2 i \eta} \right\} e^{i \alpha_{ji} t}.  \label{eq:B2}
\end{align}
Here we introduced the phenomenological parameter $\eta$ to approximately incorporate the effect of scattering. For a derivation, we refer the reader to Ref.~\onlinecite{Ward1965}, where a perturbative expression for optical rectification is derived following the perturbation theory with a damped excited-state wavefunction of Weisskopf and Wigner.\cite{Weisskopf1,Weisskopf2}  The relative signs of $i\eta$ here are equivalent to that in Eq.~8 in Ref.~\onlinecite{Buckingham2000} and with Eqs.~38 and 52 in Ref.~\onlinecite{Norman2006}.  We note that there is physical significance only in the relative signs of $\eta$ between various terms in $B_j^{(1)}$, $B_j^{(2)}$, and their conjugates, and that globally swapping $\eta \rightarrow -\eta$ in all terms doesn't change the resulting $M^{\rm IFE}$.

\section{\label{app:nonresonant} Non-doubly-resonant contributions}

The interband non-doubly resonant contribution to the IFE can be computed as,
\begin{align}
\label{eq:ndr} 
M^{\rm IFE}_{\rm ndr}
=
M^{\rm IFE}_{\rm ndr,a}
+M^{\rm IFE}_{\rm ndr,b}
+M^{\rm IFE}_{\rm ndr,c}
+M^{\rm IFE}_{\rm ndr,d}
-M^{\rm IFE}_{\rm ndr,e}
-M^{\rm IFE}_{\rm ndr,f}
-M^{\rm IFE}_{\rm ndr,g}
-M^{\rm IFE}_{\rm ndr,h}.
\end{align}
The eight contributions to $M^{\rm IFE}_{\rm ndr}$ are,
\begin{align}
M^{\rm IFE}_{\rm ndr,a}
&
=
\int_{\rm BZ} 
\frac{d^3k}{(2\pi)^3} 
\sum_n^{\rm occ} 
\sum_{N=1}^2
\sum_m^{\rm emp}
\sum_{M=1}^2
\sum_{m'}^{\rm emp}
\sum_{M'=1}^2
\left( 1-\delta_{m m'}\right)
\frac{
\braket{\phi_{nN} | V | \phi_{mM}} 
\braket{\phi_{mM} | M^{\rm spin} | \phi_{m'M'}}
\braket{\phi_{m'M'} | V^{\dagger} | \phi_{nN}} }
{
(E_m - E_n - \hbar \omega - i \eta)
(E_{m'} - E_n - \hbar \omega + i \eta)
} \label{eq:ndr-a}
\\
M^{\rm IFE}_{\rm ndr,b}
&
=
\int_{\rm BZ} 
\frac{d^3k}{(2\pi)^3} 
\sum_n^{\rm occ} 
\sum_{N=1}^2
\sum_m^{\rm emp}
\sum_{M=1}^2
\sum_{m'}^{\rm emp}
\sum_{M'=1}^2
\frac{
\braket{\phi_{nN} | V^{\dagger} | \phi_{mM}} 
\braket{\phi_{mM} | M^{\rm spin} | \phi_{m'M'}}
\braket{\phi_{m'M'} | V | \phi_{nN}} }
{
(E_m - E_n + \hbar \omega - i \eta)
(E_{m'} - E_n + \hbar \omega + i \eta)
}\label{eq:ndr-b}
\\
M^{\rm IFE}_{\rm ndr,c}
&
=
\int_{\rm BZ} 
\frac{d^3k}{(2\pi)^3} 
\sum_n^{\rm occ} 
\sum_{N=1}^2
\sum_m^{\rm emp}
\sum_{M=1}^2
\sum_{m'}^{\rm emp}
\sum_{M'=1}^2
2 {\rm Re}
\frac{
\braket{\phi_{nN} | M^{\rm spin} | \phi_{mM} }
\braket{\phi_{mM} | V | \phi_{m'M'}}
\braket{\phi_{m'M'} | V^{\dagger} | \phi_{nN}} }
{
(E_m - E_n + 2 i \eta)
(E_{m'} - E_n - \hbar \omega + i \eta)
}\label{eq:ndr-c}
\\
M^{\rm IFE}_{\rm ndr,d}
&
=
\int_{\rm BZ} 
\frac{d^3k}{(2\pi)^3} 
\sum_n^{\rm occ} 
\sum_{N=1}^2
\sum_m^{\rm emp}
\sum_{M=1}^2
\sum_{m'}^{\rm emp}
\sum_{M'=1}^2
2 {\rm Re}
\frac{
\braket{\phi_{nN} | M^{\rm spin} | \phi_{mM} }
\braket{\phi_{mM} | V^{\dagger} | \phi_{m'M'}}
\braket{\phi_{m'M'} | V | \phi_{nN}} }
{
(E_m - E_n + 2 i \eta)
(E_{m'} - E_n + \hbar \omega + i \eta)
} \label{eq:ndr-d}
\\
M^{\rm IFE}_{\rm ndr,e}
&
=
\int_{\rm BZ} 
\frac{d^3k}{(2\pi)^3} 
\sum_n^{\rm occ} 
\sum_{N=1}^2
\sum_{n'}^{\rm occ}
\sum_{N'=1}^2
\sum_m^{\rm emp}
\sum_{M=1}^2
\left( 1-\delta_{n n'}\right)
\frac{
\braket{\phi_{nN} | V | \phi_{mM}} 
\braket{\phi_{mM} | V^{\dagger} | \phi_{n'N'}}
\braket{\phi_{n'N'} | M^{\rm spin} | \phi_{nN}} }
{
(E_m - E_n - \hbar \omega - i \eta)
(E_m - E_{n'} - \hbar \omega + i \eta)
} \label{eq:ndr-e}
\\
M^{\rm IFE}_{\rm ndr,f}
&
=
\int_{\rm BZ} 
\frac{d^3k}{(2\pi)^3} 
\sum_n^{\rm occ} 
\sum_{N=1}^2
\sum_{n'}^{\rm occ}
\sum_{N'=1}^2
\sum_m^{\rm emp}
\sum_{M=1}^2
\frac{
\braket{\phi_{nN} | V^{\dagger} | \phi_{mM}} 
\braket{\phi_{mM} | V | \phi_{n'N'}}
\braket{\phi_{n'N'} | M^{\rm spin} | \phi_{nN}} }
{
(E_m - E_n + \hbar \omega - i \eta)
(E_m - E_{n'} + \hbar \omega + i \eta)
}\label{eq:ndr-f}
\\
M^{\rm IFE}_{\rm ndr,g}
&
=
\int_{\rm BZ} 
\frac{d^3k}{(2\pi)^3} 
\sum_n^{\rm occ} 
\sum_{N=1}^2
\sum_{n'}^{\rm occ}
\sum_{N'=1}^2
\sum_m^{\rm emp}
\sum_{M=1}^2
2 {\rm Re}
\frac{
\braket{\phi_{nN} | M^{\rm spin} | \phi_{mM} }
\braket{\phi_{mM} | V^{\dagger} | \phi_{n'N'}}
\braket{\phi_{n'N'} | V | \phi_{nN}} }
{
(E_m - E_n + 2 i \eta)
(E_m - E_{n'} - \hbar \omega + i \eta)
} \label{eq:ndr-g}
\\
M^{\rm IFE}_{\rm ndr,h}
&
=
\int_{\rm BZ} 
\frac{d^3k}{(2\pi)^3} 
\sum_n^{\rm occ} 
\sum_{N=1}^2
\sum_{n'}^{\rm occ}
\sum_{N'=1}^2
\sum_m^{\rm emp}
\sum_{M=1}^2
2 {\rm Re}
\frac{
\braket{\phi_{nN} | M^{\rm spin} | \phi_{mM} }
\braket{\phi_{mM} | V | \phi_{n'N'}}
\braket{\phi_{n'N'} | V^{\dagger} | \phi_{nN}} }
{
(E_m - E_n + 2 i \eta)
(E_m - E_{n'} + \hbar \omega + i \eta)
}. \label{eq:ndr-h}
\end{align}
\end{widetext}

\bibliography{pap}

\end{document}